# A FIBROUS SOLID ELECTROLYTE FOR LITHIUM-ION BATTERIES

Salim EROL[1]

[1]Department of Chemical Engineering, Eskisehir Osmangazi University, Eskişehir 26040, Turkey

esalim@ogu.edu.tr

**Abstract:** The new methodology proposed in this work uses oxide filaments/fibers instead of a dense thin film electrolyte, allowing: (i) straightforward fabrication without the use of costly thin film production techniques, (ii) minor changes in the existing Li-ion battery production lines, and (iii) better mechanical stability and electrochemical performance. $Li_7La_3Zr_2O_{12}$ (LLZO) fibers as an electrolyte were prepared using the electrospinning technique. The crystallographic phase and morphology of the products was studied X-ray diffraction (XRD) patterns and scanning electron microscopy (SEM), respectively. The results indicated that the produced electrolyte would be a good alternative of all-solid-state batteries for electric vehicles in terms of battery safety and performance.

INTRODUCTION

Lithium-ion batteries are currently the dominant power source for portable electronic devices because of their high energy density and efficiency. Recently, there is a strong growing interest to develop large-scale Li-ion batteries, particularly for electric vehicles. Developing batteries with high energy density, safe operation, low cost, and long life is vital for successful implementation of Li-ion batteries in automotive industry. All-solid-state Li-ion batteries have been shown to significantly improve volumetric energy density, safety, and cycle life; however, their real-world applications are hindered by the complex and costly manufacturing processes. Recognizing this important need, the main objective of this study is to examine a new scalable and inexpensive technique for fabrication of an electrolyte for all-solid-state batteries.

Conventional Li-ion batteries typically consist of a positive electrode (cathode), a negative electrode (anode), and a liquid organic electrolyte. During charge/discharge cycles Li ions migrate through the liquid electrolyte between the two electrodes. Large-scale Li-ion batteries face many challenges in terms of safety, cost, and cycle life. Many of these challenges are associated with the use flammable liquid organic electrolytes. The liquid electrolyte can leak through the containment walls of the battery and catch on fire. The liquid electrolyte cannot provide a separation between the cathode and anode; therefor, polymeric separators need to be used in order to prevent the battery from a potential short-circuit. The use of flammable organic liquid electrolytes and polymeric separators induce serious safety problems including the risk of fire and explosion. Besides these safety concerns, the complex packaging remarkably increases the manufacturing cost, and lowers the energy density by increasing the weight and volume of the battery packs. In addition, side reactions and poor thermal stability of the liquid electrolyte result in a capacity fade, which shortens the battery cycle life. [1]

In all-solid-state Li-ion batteries, the components are all made up of solid phase. The nonflammability of the solid electrolyte overcomes the safety problem of the conventional Li-ion batteries. Since solid electrolytes neither leak nor vaporize easily, all of the cell components can be directly stacked in one container, instead of multiple containers as in the liquid-containing cells. Therefore, as shown in Fig. 1a, the battery size reduces, and thus the volumetric energy density significantly improves up to 3 times. Moreover, solid electrolytes physically separate the cathode from the anode, eliminating the need for separators. In addition, solid electrolytes have a wider range of operating potentials (over 5 V) and operating temperatures. They can be paired with certain high-voltage and high-capacity electrodes that are generally difficult to combine with liquid electrolytes. The good examples of these electrodes are lithium metal anodes and sulfur cathodes. Another feature of solid electrolytes is their single-ion conduction that suppresses side reactions and leads to a long battery life. As a result of unity transference number indicating that Li ions are the sole charge carriers, there is no concentration gradient inside the operating cells; thus, the cell over-potential is negligible. Furthermore, the capacity fade caused by the continuous decomposition of the electrode surface in contact with liquid is avoided. [2-4]

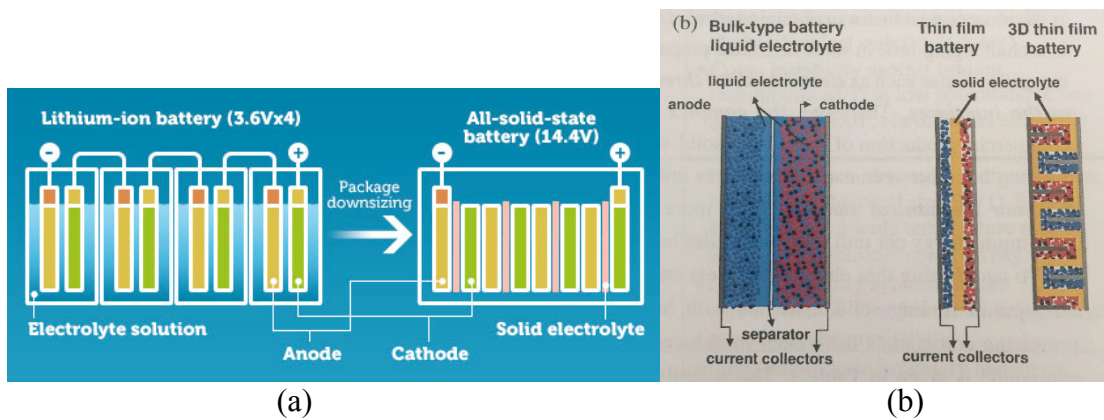

(a)                                (b)

**Fig. 1.** (a) Improved packaging efficiency by direct stacking of the all-solid-state cells instead of using multiple containers in liquid-containing cells, (b) A comparison between the Li-ion accessibility in bulk-type liquid-containing cells, and 2D and 3D all-solid-state thin film batteries.

In summary, solid electrolytes improve safety, offer higher volumetric and gravimetric energy densities, and potentially lower the cost of the battery by increasing the battery life, decreasing the dead-space in battery packs, and simplifying the packaging.

Li-ion batteries are the preferred system for large-scale energy storage, particularly in automobiles. Most common rechargeable Li-ion batteries are made using organic liquid electrolytes, which is currently the major constraint for design and production of these batteries for large-scale applications. The liquid electrolyte can leak through the containment walls of the battery and can catch fire or cause an explosion due to the flammability of lithium exposing to air. The replacement of the organic liquid electrolyte with a more reliable inorganic solid electrolyte in all-solid-state Li-ion batteries simplifies the design of these batteries and improves their durability. Compared to liquid electrolytes, nonflammable solid electrolytes significantly improve the safety of the batteries. Moreover, direct stacking of all-solid-state cells in one package yields a high operating voltage in a smaller volume and provides higher energy density. In order to achieve the potential advantages of solid-state batteries and to commercialize this class of batteries, thin films of the solid electrolytes were developed to provide low resistance for Li-ion transfer. Current vacuum-based deposition techniques such as pulsed-laser deposition, atomic layer deposition, and magnetron sputtering that are used in laboratory environments and electronics industry are too costly and therefore do not allow for scale-up of the process and commercial production of large-scale batteries. Furthermore, a good contact between the solid electrolyte and electrode materials cannot be readily obtained, making it difficult to fabricate an all-solid-state battery.

The solid electrolyte is the most critical component in all-solid-state batteries. It should exhibit a high Li-ion conductivity to facilitate shuttling of the Li ions between the two electrodes. Moreover, it should have a low electronic conductivity to electronically separate the electrodes and to lower the self-discharge. The chemical stability in contact with other cell components and under the operating potential and temperature is another important property that a solid electrolyte should possess.

Inorganic solids, particularly oxides, usually have a low ionic conductivity. The Li-ion conductivity of lithium phosphorus oxynitride (LIPON), which is $10^{-6}$ S cm$^{-1}$ with a transference number of 1 [5,6], known as a good Li-ion conductor, is still 4 orders of magnitude lower than the commercial liquid electrolytes, which have $10^{-2}$ S cm$^{-1}$ conductivity with a transference number of <0.5 [7]. Therefore, thin films (<200 nm) of the solid electrolytes were developed to provide low resistance for Li-ion transfer. [5] In addition to the low ionic conductivity, the solid electrolyte, unlike liquids, cannot penetrate through the pores inside the porous electrodes. Thus, the contact area between the electrodes and the solid electrolyte is very limited to, as demonstrated in Fig. 1b. The small contact area limits the accessibility of ions to the electrode active sites, resulting in low utilization of the electrode materials. Various designs of 3D thin film batteries (Fig. 1b), by providing a larger macroscopic contact area, could partly address this issue. [8-12] Currently, the thin film batteries are fabricated using costly vacuum-based techniques, such as pulsed-layer deposition, atomic layer deposition, and magnetron sputtering. [6,13-16] The concentration of chemicals and point defects is the most critical factor in obtaining ideal electrical properties, while an exact stoichiometric control is a challenging task in most thin film preparation techniques. Scaling down the film thickness also magnifies issues such as electrical short through pinholes in the film and poor electrical conductivity at the interfaces. Therefore, the complex design of thin film batteries limits the scale-up and commercial production of large-scale solid-state batteries. As film thickness is increased, maintaining a strong bond between each of the films and the supporting substrate becomes more difficult. Thus, the poor mechanical stability, even more than the electrical resistance, dictates the design and maximum energy per unit area of the film batteries. [17]

There have been extensive attempts in finding solid materials with Li-ion conductivity to avoid using thin electrolyte layers and costly vacuum technologies. As a result, a number of promising compounds have been introduced. A summary of advantages and disadvantages of each compound is depicted in Table 1. The recently reported sulfide compounds [3] exhibit significantly a high Li-ion conductivity ($10^{-2}$ S cm$^{-1}$) that is comparable to that of liquid electrolytes. However, the instability of the sulfide glasses in direct contact with lithium and the hygroscopic feature of sulfide compounds can constraint their application.

**Table 1.** Examples of Li-ion conducting inorganic compounds considered as electrolyte material in all-solid-state Li-ion batteries.

| Groups of Materials | | Example(s): Li-ion conductivity | Advantages & Disadvantages |
|---|---|---|---|
| Sulfides | Sulfide glasses | LiI-Li$_2$S-X (X: P$_2$S$_5$, B$_2$S$_3$, SiS$_2$): 2×10$^{-3}$ S cm$^{-1}$ [18-20] | Pros: high ionic conductivity in the absence of transition metal oxide (wide electrochemical window), low grain boundary resistance even in a cold-pressed pellet (convenient for the bulk-type batteries) Cons: low chemical stability in air, synthesis under controlled atmosphere, instability against Li metal, high resistance at the LiCoO$_2$/sulfide interface |
| | Crystalline sulfides | Li$_{10}$GeP$_2$S$_{12}$ (Thio-LISICON): 1.2×10$^{-2}$ S cm$^{-1}$ [21] | |
| | Glass-ceramic sulfides | Li$_2$S-P$_2$S$_5$: 3.2×10$^{-3}$ S.cm$^{-1}$ [22,23] | |
| Oxysalts | NASICON-type | Al-doped LiTi$_2$(PO$_4$)$_3$ (LTP): 3×10$^{-3}$ S cm$^{-1}$ [5] | Pros: stability in air and water, high electrochemical stability (1.8-6 V) Cons: instability against Li (reduction of Ti at low potentials) [24,25] |
| | γ-Li$_3$PO$_4$-type | Amorphous lithium phosphorus oxynitride (LIPON): 2×10$^{-6}$ S cm$^{-1}$ [26] | Pros: wide electrochemical potential window (0-5.5 V) Cons: sufficient conductance and mechanical stability only when thinner than 2 $\mu$m [6,27] |
| Oxides | Perovskite-type | Li$_{3x}$La$_{(2/3)-x}$TiO$_3$ (LLTO) with modified grain boundaries: 1×10$^{-4}$ S cm$^{-1}$ [28,29] | Pros: stability in dry and wet atmosphere, high electrochemical stability (1.8-8 V), stability over a wide temperature range, negligible electronic conductivity Cons: instability against Li (reduction of Ti at low potentials), high grain boundary resistance |
| | Garnet-type | Li$_7$La$_3$Zr$_2$O$_{12}$ (LLZO): 8.7×10$^{-4}$ S cm$^{-1}$ [30] | Pros: low grain boundary resistance, excellent chemical stability (stability against Li), wide electrochemical potential window (0-9 V) [31] |

EXPERIMENTAL

Fabrication of the electrolyte in the form of fiber, whisker, and/or filament was performed using the electrospinning technique. This technique is widely used for fabricating fibers from a wide range of materials, including oxysalts, oxides, and sulfides, with diameters ranging from several nanometers to micrometers and lengths up to several millimeters. The process is based on the unidirectional elongation of a spinnable viscoelastic solution by considering various parameters involved in the electrospinning process. [32] For the experiments, a solution containing the stoichiometric amount of oxide precursors (lithium acetate, zirconium, and lanthanum nitrate hexahydrate), acetic acid, and granular polyvinylpyrrolidone (PVP) dissolved in ethanol were drawn into a 15-gauge syringe with a blunt-tip stainless steel needle that was placed 15 cm from the collector. The ceramic precursor to PVP ratio was 1:2. During electrospinning, a positive voltage of 20 kV was applied to the needle, and the solution feed rate of 0.1 mL h$^{-1}$ was maintained using a syringe pump. The schematics of the electrospinning process are illustrated in Fig. 2. The obtained inorganic-organic composite fibers were collected on an ashless filter paper. The paper was placed on the metal collector, which was an aluminum foil. The fibers and the ashless paper removed from the collector after the electrospinning process and dried at 80 ºC for 12 h. Then, the spun fiber was calcined at 900 ºC to obtain the final oxide fiber. The crystallographic phase and morphology of the calcined fibers was studied X-ray diffraction (XRD) patterns and microscopic images. Various parameters that can heavily influence the morphology of the fibers, including processing parameters (the applied electric potential, the solution rate, the diameter of the syringe and the needle, the syringe tip-to-collector distance, and the calcination temperature) and solution parameters (viscosity, pH, and the precursor to PVP ratio) were all investigated by repeating the experiments several times, and the parameter values were optimized as mentioned above.

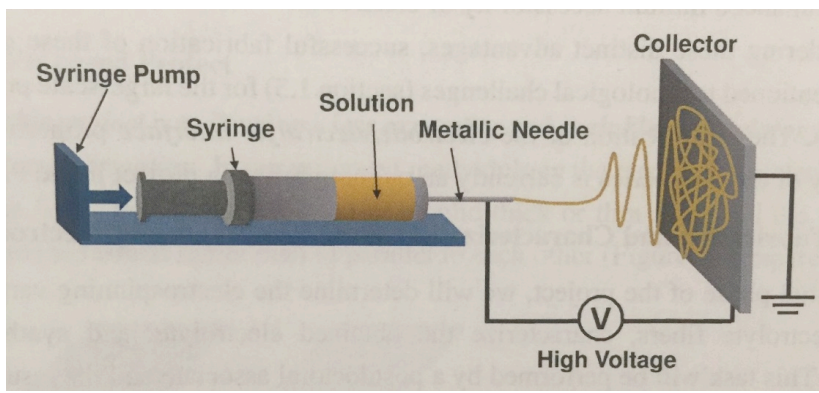

**Fig. 2.** Schematic representation of electrospinning process.

## RESULTS AND DISCUSSION

Microscopic images of spun LLZO fibers before calcination and after calcination are shown in Fig. 3. The images reveal that the fibrous structure was maintained an even improved after the calcination process.

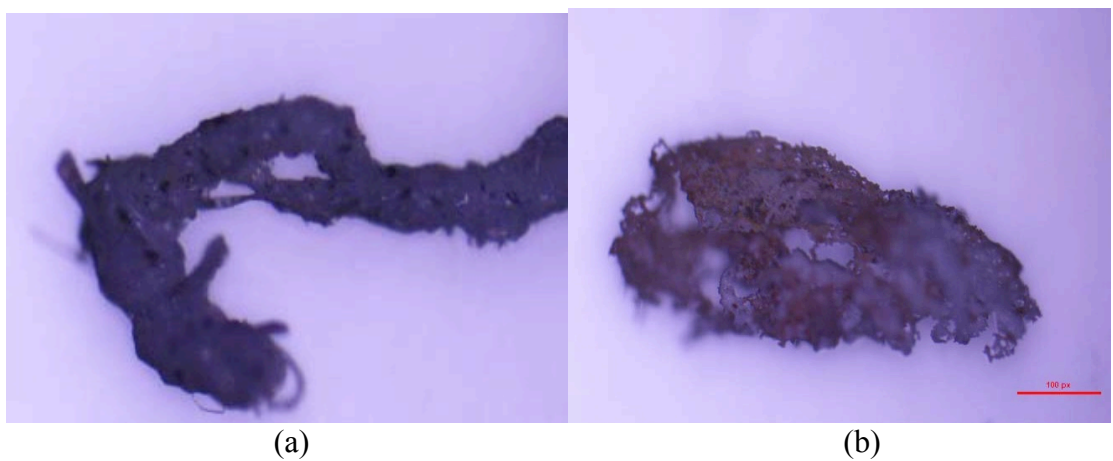

(a)                                  (b)

**Fig. 3.** Microscopic images of spun LLZO fibers (a) before calcination and (b) after calcination. The scale bar in (b) is 100 px (40 $\mu$m), and the scaling is same for both images.

The XRD patterns of the calcined LLZO sample are demonstrated in Fig. 4. This specific pattern shows that the LLZO obtained in this study is in cubic structure. The cubic LLZO has higher conductivity than that of tetragonal.

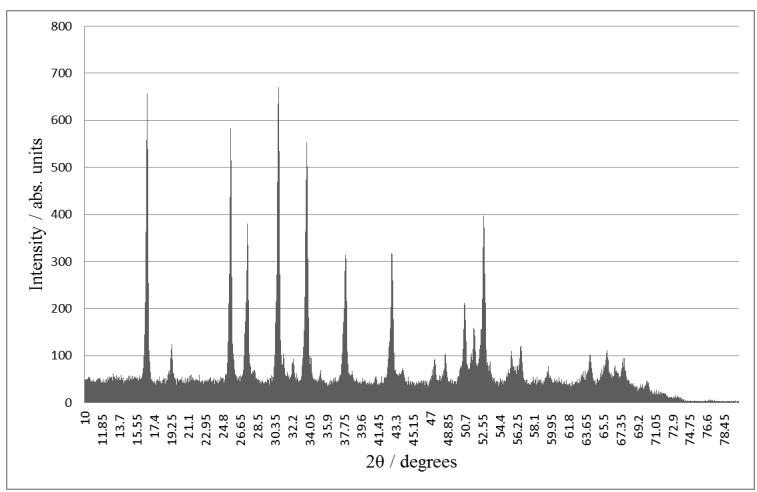

**Fig. 4.** X-ray diffraction patterns for LLZO sample

The microscopic images and XRD results suggest that the fabricated LLZO fibers are ready to be used as an electrolyte for all-solid-state Li-ion batteries.

CONCLUSION AND FUTURE WORK

Electrospinning is an easy and inexpensive method to form electrolyte fibers, filaments, or whiskers. The fibrous LLZO electrolyte fabricated using the electrospinning technique might be an appropriate electrolyte of all-solid-state battery for electric vehicles in terms of battery safety and performance.

A high capacity all-solid-state Li-ion battery using this type electrolyte is proposed to present in a subsequent paper. In this proposed cell configuration, the fibrous electrolyte will be placed on an insulating layer; then a very narrow insulating glue will be applied in the center. The prepared electrode slurries will be spread on the both sides of the insulating glue, and current collectors (foil or mesh) will be attached.